\documentclass[draft]{mn2e}

\usepackage{psfig}
\def\gs{\mathrel{\raise0.35ex\hbox{$\scriptstyle >$}\kern-0.6em
\lower0.40ex\hbox{{$\scriptstyle \sim$}}}}
\def\ls{\mathrel{\raise0.35ex\hbox{$\scriptstyle <$}\kern-0.6em
\lower0.40ex\hbox{{$\scriptstyle \sim$}}}}
\def\ls{\mathrel{\hbox{\rlap{\hbox{\lower4pt\hbox{$\sim$}}}\hbox{$<$}}}}
\def\gs{\mathrel{\hbox{\rlap{\hbox{\lower4pt\hbox{$\sim$}}}\hbox{$>$}}}}

\title[How typical is the Coma cluster?]
      {How typical is the Coma cluster?}
\author[K.\,A.\ Pimbblet, S.\,J.\ Penny, R.\,L.\ Davies ]
       {Kevin A.\ Pimbblet$^{1,2,3,4}$\thanks{email: Kevin.Pimbblet@monash.edu},
Samantha J.\ Penny$^{2,3}$,
Roger L.\ Davies$^{4}$
        \vspace*{1mm}\\
$^{1}$Department of Physics and Mathematics, University of Hull, Cottingham Road, Hull, HU6 7RX, UK\\
$^{2}$School of Physics, Monash University, Clayton, Victoria 3800, Australia\\
$^{3}$Monash Centre for Astrophysics (MoCA), Monash University, Clayton, Victoria 3800, Australia\\
$^{4}$Department of Physics, University of Oxford, Denys Wilkinson Building, Keble Road, Oxford OX1 3RH, UK\\
}

\date{Accepted ... ; Received ... ; in original ...}

\pagerange{000--000}

\begin{document}

\maketitle

\begin{abstract}
Coma is frequently used as the archetype $z\sim0$ galaxy cluster to compare
higher redshift work against. It is not clear, however, how representative the 
Coma cluster is for galaxy clusters of its mass or X-ray luminosity,
and significantly: recent works have suggested
that the galaxy population of Coma may be in some ways anomolous.
In this work, we present a comparison of Coma to an X-ray selected control
sample of clusters.
We show that although Coma is typical against the control
sample in terms of its internal kinematics (substructure and velocity dispersion profile), 
it has a significantly high ($\sim3\sigma$) X-ray
temperature set against clusters of comparable mass. 
By de-redshifting our control sample cluster galaxies star-formation rates using a fit 
to the galaxy main sequence evolution at $z<0.1$, we determine that the typical
star-formation rate of Coma galaxies as a function of mass is higher
than for galaxies in our control sample at a confidence level of $>99$ per cent. 
One way to alleviate this discrepency and bring Coma
in-line with the control sample would be to have the distance to
Coma to be slightly lower, perhaps through a non-negligible peculiar 
velocity with respect to the Hubble expansion,
but we do not regard this as likely given precision
measurements using a variety of approaches.
Therefore in summary, we urge caution in using Coma
as a $z\sim0$ baseline cluster in galaxy evolution studies.

\end{abstract}

\begin{keywords}
galaxies: clusters: individual: Coma cluster ---
galaxies: clusters: general ---
galaxies: evolution ---
X-rays: galaxies: clusters
\end{keywords}

\section{Introduction}
Clusters of galaxies span a wide range of physical conditions and internal
configurations. At the high mass end of their mass distribution ($\sim 10^{15}$ 
solar masses), clusters may
contain several thousand member galaxies that are orbiting with velocity
dispersions over 1000 kms$^{-1}$ (cf.\ Pimbblet et al.\ 2006; Ebeling et al.\ 2010).  
They are also rare celestial objects: they form from the gravitational collapse
of extremely large perturbations within the primorial density field 
(e.g.\ Zel'Dovich 1970; Doroshkevich \& Shandarin 1978)
and continue
to grow at all epochs through the accretion of fresh material; a large fraction
of galaxies being funnelled directly to them 
through the filaments of the cosmic web (Pimbblet, Drinkwater \& Hawkrigg 2004).
From the point of view of studying galaxy evolution, clusters of galaxies
offer excellent test-beds as they contain a range conditions from their outskirts
(which may contain filaments and under-dense `void' regions that galaxies are
being accreted from) through to high density cores that contain a dense, hot 
($10^7$--$10^8$ K) X-ray emitting gas that is capable of stripping an infalling
galaxy of its own star-forming gas (Gunn \& Gott 1972; Cayatte et al.\ 1990;
Quilis et al.\ 2000; Boselli \& Gavazzi 2006).
Indeed, galaxies that are located at the centre of clusters (or high density regions
of the Universe) have long been noted to possess systematically different
properties (star-formation rates; colours; morphologies; masses) to those in low
density regions (e.g.\ Dressler 1980; Lewis et al.\ 2002; G{\'o}mez et al.\ 2003; 
Baldry et al.\ 2006;
Bamford et al.\ 2009; Wilman \& Erwin 2012; amongst many others). To address 
questions concerning the evolution of galaxies within these structures, samples
of self-similar structures (and/or their likely progenitors) 
need to be assembled across cosmic time.

The Coma cluster (also known as Abell 1656 in the catalogue of Abell 1958) is 
the closest galaxy cluster of its mass (recently derived to be $1.8 \times 10^{15}$
solar masses through a weak lensing analysis by Kubo et al.\ 2007) to us.  
This has lead to Coma being extensively used as a redshift $z\approx0$ baseline to 
compare higher redshift galaxy clusters to (e.g.\ 
Bahcall 1972;
Mellier et al.\ 1988;
Stanford, Eisenhardt \& Dickinson 1995;
Smith, Driver \& Phillipps 1997;
Kodama et al.\ 1998;
J{\o}rgensen et al.\ 1999;
Jones, Smail \& Couch 2000;
Kodama \& Bower 2001;
van Dokkum et al.\ 2001;
La Barbera et al.\ 2002;
Rusin et al.\ 2003;
De Lucia et al.\ 2004; 
Ellis \& Jones 2004;
Poggianti et al.\ 2004;
Fritz et al.\ 2005;
Holden et al.\ 2005
De Lucia et al.\ 2007; 
Moran et al.\ 2007;
van Dokkum \& van der Marel 2007;
D'Onofrio et al.\ 2008;
Giard et al.\ 2008;
Ascaso et al.\ 2009; 
Bai et al.\ 2009; 
Lah et al.\ 2009; Stott et al.\ 2009).

Yet it is not clear how representative (or `typical') the Coma cluster
is for clusters of its mass. To illustrate this point, we note two
recent examples.
Stott et al.\ (2009) present an analysis of how the slope of the
colour-magnitude relation (Visvanathan \& Sandage 1977) of clusters
varies with redshift. They find that the rest-frame slope evolves according to
$(1+z)^{1.77}$ (see their Fig.~7).  Yet, the slope for the Coma cluster lies
at least $2\sigma$ away (steeper) from this relationship and its absolute value is much more
in-line with what might be expected of a $z\sim0.3$ cluster (Stott et al.\ 2009).  
Indeed, the inclusion of Coma
pulls their power law fit upward at the low redshift end as it is the only point
they consider at $z<0.08$.  Stott et al.\ (2009) attribute this mildly unusual slope
to a lower than average
dwarf-to-giant ratio along its red sequence (Stott et al.\ 2007)
that suggests it is still undergoing significant faint end evolution.
Whilst Stott et al.'s result is likely
not a statistically significant issue, other studies yield stronger issues with 
the use of Coma as a $z\approx0$ baseline.
Pertinent to this is the second example of
Ascaso et al.\ (2009; see also Ascaso et al.\ 2008) who measure
the structural properties (e.g.\ surface brightness profiles and quantitative galaxy
morphologies)
for a sample galaxies taken from 5 clusters at $0.18<z<0.25$ and compare
them to Coma (using data from Aguerri et al.\ 2004). 
They find that the scales of the discs of late-type galaxies 
in the high redshift clusters are
significantly different to Coma.  They offer two conclusions: either spiral galaxies
have undergone a remarkable and very strong 
evolution over the past 2.5 Gyr, or `Coma is in some way
anomalous' (Ascaso et al.\ 2009).

Much earlier studies that concentrate on Coma itself describe the cluster
as `rich', `regular' and (or) `relaxed' (e.g.\ Kent \& Gunn 1982 retain the
assumption of the cluster being in equilibrium; see also 
Noonan 1961;
Omer, Page \& Wilson 1965 and references therein).
Evidence subsequently accumulated that Coma was anything but a local archetype
for relaxed and regular clusters: 
Henriksen \& Mushotzky (1986) used X-ray observations to invalidate the
assumption of an isothermal sphere (see also Johnson et al.\ 1979; 
Briel, Henry \& Boehringer 1992; White, Briel \& Henry 1993;
Vikhlinin, Forman \& Jones 1997;
Neumann et al.\ 2003); 
the cluster contains multiple D class galaxies (Beers \& Geller 1983);
and importantly the velocity distribution of the galaxy members 
themselves revealed substructure (Fitchett \& Webster 1987; 
Merritt 1987;
Mellier et al.\ 1988;
Colless \& Dunn 1996; Gambera et al.\ 1997;
Edwards et al.\ 2002; Adami et al.\ 2009;
see also Conselice \& Gallagher 1998).

The central thesis of this work is to present a
novel investigation of how typical the Coma cluster is
in three well-defined and 
distinct ways that are well-used in the literature.
This comprises: (i) an investigation in to the X-ray properties 
(particularly temperature and luminosity)
of Coma in comparison to analogue clusters; (ii) a consideration
of how kinematically perturbed or relaxed analogue clusters are to Coma;
(iii) a determination of how `active' -- in the sense of star-formation --
the galaxies that make up analogous clusters are compared to Coma.
The format for this work is as follows.  
In Section~2, we describe the creation of a set of control clusters
that are analogous to Coma in mass from available SDSS and X-ray data.
We examine the X-ray properties of Coma in comparison to the control
sample and an extended sample in Section~3. Section~4 deals with the
kinematics of the galaxies contained in the clusters and in Section~5,
we examine the star-formation rates of the constituent galaxies in Coma
and the control sample. Our results are summarized in Section~6.
Throughout this work, we have used the Spergel et al.\ (2007) standard, flat 
cosmology in which 
$\Omega_M = 0.238$, $\Omega_{\Lambda}=0.762$ and $H_0=73$ km s$^{-1}$ Mpc$^{-1}$.

\section{Data}

%
%
\begin{table*}
\begin{center}
\caption{The
sample of clusters used in this work.
The coordinates specify the Vizier position of the cluster.
The X-ray luminosities in the 0.1--2.4 keV band
($L_X$) and temperatures ($T_X$)
are sourced from BAX (Sadat et al.\ 2004) which is a compilation
of X-ray data deriving from many diverse literature sources. We cite
the sources of these values below the table using brackets next to each value.
The virial radius ($R_{virial}$) is computed
from $\sigma_{cz}$; see text for details. Bautz-Morgan (B-M) types have been
sourced from NED except for the Zwicky clusters which we have determined ourselves.
\hfil}
\begin{tabular}{lllllllll}
\noalign{\medskip}
\hline
Name  & RA      & Dec     & B-M  & $L_X$                           & $T_X$ & $\overline{cz}$ & $\sigma_{cz}$  & Adopted  \\
      & (J2000) & (J2000) & type  & ($\times 10^{44}$ erg s$^{-1}$) & (keV) & (km\,s$^{-1}$)  & (km\,s$^{-1}$) & $R_{200}$ (Mpc)      \\
\hline
Abell~85 & 00 41 38 & $-$09 20 33 & I  & 9.41 (i) & 6.45$^{+0.10}_{-0.10}$ (a) & 16537$\pm$58 & 898$^{+45}_{-39}$ & 1.80 \\
Abell~119 & 00 56 21 & $-$01 15 47 & II-III  & 3.30 (i) & 5.62$^{+0.12}_{-0.12}$ (b) & 13381$\pm$56 & 917$^{+43}_{-37}$ & 1.83 \\
Abell~660 & 08 25 22 & +36 50 16 & III &  4.39 (ii) & N/A  & 19647$\pm$109 & 681$^{+94}_{-66}$ & 1.36 \\
ZwCl~1215.1+0400 & 12 17 41 & +03 39 32 & II  &  5.17 (i) & 6.54$^{+0.21}_{-0.21}$ (a) & 23109$\pm$54 & 867$^{+41}_{-36}$ & 1.73 \\
Abell~1775 & 13 41 56 & +26 21 53 & I & 2.80 (i) & 3.66$^{+0.21}_{-0.12}$ (c) & 21593$\pm$113 & 1637$^{+86}_{-74}$ & 3.27 \\
Abell~1795 & 13 49 01 & +26 35 07 & I & 10.26 (i) & 6.12$^{+0.05}_{-0.05}$ (a) & 18773$\pm$55 & 719$^{+42}_{-36}$ & 1.44 \\
Abell~1800 & 13 49 41 & +28 04 08 & II & 2.89 (i) & 4.14$^{+0.09}_{-0.09}$ (d) & 22915$\pm$80 & 1018$^{+61}_{-52}$ & 2.04 \\
ZwCl~1518.8+0747 & 15 21 52 & +07 42 31 & I & 2.78 (i) & 3.45$^{+0.08}_{-0.06}$ (c) & 13408$\pm$88 & 1064$^{+68}_{-57}$ & 2.13 \\
Abell~2061 & 15 21 15 & +30 39 17 & III & 4.85 (iii) & 4.52$^{+0.10}_{-0.10}$ (d) & 23083$\pm$52 & 841$^{+39}_{-35}$ & 1.63 \\
Abell~2065 & 15 22 43 & +27 43 21 & III & 5.55 (i) & 5.44$^{+0.09}_{-0.09}$ (a) & 22213$\pm$100 & 1887$^{+75}_{-67}$ & 3.77 \\
Abell~2199 & 16 28 39 & +39 33 06 & I & 4.09 (i) & 3.99$^{+0.10}_{-0.10}$ (a) & 9176$\pm$44 & 761$^{+33}_{-29}$ & 1.52 \\
Abell~2255 & 17 12 31 & +64 05 33 & II-III & 5.54 (i) & 5.92$^{+0.24}_{-0.16}$ (c) & 24071$\pm$82 & 1223$^{+62}_{-54}$ & 2.45 \\
\hline
Coma & 12 59 49 & +27 58 50 & II & 7.77 (i) & 8.25$^{+0.10}_{-0.10}$ (e) & 7166$\pm$54 & 1639$^{+40}_{-37}$ & 3.28 \\
\hline
\multicolumn{3}{l}{(i) Reiprich \& B{\"o}hringer (2002)} & \multicolumn{3}{l}{(a) Vikhlinin et al.\ (2009)} \\
\multicolumn{3}{l}{(ii) Popesso et al.\ (2007a) } & \multicolumn{3}{l}{(b) Henry (2004)} \\
\multicolumn{3}{l}{(iii) Marini et al.\ (2004) } & \multicolumn{3}{l}{(c) Ikebe et al.\ (2002)} \\
\multicolumn{3}{l}{ } & \multicolumn{3}{l}{(d) Shang \& Scharf (2009)} \\
\multicolumn{3}{l}{ } & \multicolumn{3}{l}{(e) Arnaud et al.\ (2001)} \\

\label{tab:sample}
\end{tabular}
\end{center}
\end{table*}

We use two sets of data in this work, both taken from SDSS
Data Release 7 (Abazajian et al.\ 2009). The first set of data is
for the Coma cluster itself, whilst the second set (the control sample)
consists of SDSS clusters that possess comparable X-ray 
luminosity (an observational proxy for mass since it
originates from thermal Bremsstrahlung of the hot intra-cluster gas) 
to Coma. We make use of the SDSS value-added catalogues throughout this
work, which includes star-formation rates (Brinchmann et al.\ 2004)
and masses (see www.mpa-garching.mpg.de/SDSS/DR7/Data/stellarmass.html).

To create the control sample, we note that 
Coma has an X-ray luminosity of $L_X = 7.77\times10^{44}$ erg~s$^{-1}$ 
measured in the 0.1--2.4 keV band
(Reiprich \& B{\"o}hringer 2002).  This level of emission is
comparable with some of the most massive clusters in the Universe
(cf.\ Ebeling et al.\ 2001; Pimbblet et al.\ 2001).
We therefore would like to select clusters with comparable
$L_X$ in the 0.1--2.4 keV band, but balance this with a need to have a sufficiently
large control sample to contrast Coma against.  We therefore
select clusters within $5 \times10^{44}$ ergs$^{-1}$ of Coma's 
X-ray luminosity. Since X-ray luminosity can predict cluster
mass with an accuracy of $>$50 per cent, 
such a range is likely to correspond to no more than a factor
of 2 range in mass from this $L_X$ selection (Popesso et al.\ 2005).
Secondly, we would like to select galaxy clusters to be at
a comparable stage in their evolution as Coma.  
We firstly note that Kodama \& Smail (2001) suggest the time-scale
for galaxy morphological transformation within clusters
may be as short as 1 Gyr if gas starvation effects are
strong (see also 
Bekki, Couch \& Shioya 2002;
Moran et al.\ 2006; 
Tonnesen et al.\ 2007;
Boselli et al.\ 2008).
Therefore we wish to select clusters within a $<$1 Gyr look-back 
of Coma.
This corresponds
to a maximum redshift of $z\sim0.08$ to select our clusters from.

We use the Base de Donnees Amas de Galaxies X
(BAX) X-Ray Clusters Database (Sadat et al.\ 2004)
to select clusters from using the above criteria.
This yields a total of 47 clusters.  
Of these, one is Coma and
a further 13 (30 per cent) are within the spatial limits of SDSS
-- this criteria of being within the observational bounds of SDSS
is only applied after the X-ray selection within BAX.
We detail the global properties
of these clusters in Table~\ref{tab:sample}, alongside Coma.
We note that the clusters in the control sample 
have a mean $L_X = 5.1 \pm 2.4 \times10^{44}$ erg~s$^{-1}$ -- only $\sim 1\sigma$ less
than Coma's.

From this sample, we exclude NRGB045
on the grounds that it has an anomolously low $T_X$ value
(0.83 keV).
This is due the 
NRGB045 being more akin to a group than a cluster.  Indeed,
recent work by Stott et al.\ (2012) suggests that any
galaxy grouping with $T_X<2$ keV would physically be
considered a group rather than a bona fide cluster.  
The exclusion of NRGB045 from our subsequent analysis
leaves us with 12 clusters in the control sample.

For each of the clusters in our control sample, we download all galaxies within
1 deg of the BAX-specified cluster centres from SDSS.
For each cluster, we derive new estimates of their mean
recession velocity ($\overline{cz}$) and velocity dispersion ($\sigma_{cz}$)
from the `gapping' technique of Zabludoff, Huchra \& Geller (1990; 1993) which
iteratively eliminates any galaxy from the computation of $\overline{cz}$ 
that is deviant by more than $3\sigma_{cz}$ from $\overline{cz}$.
Errors on $\sigma_{cz}$ are generated following Denese, de Zotti, G. \& di Tullio (1980).
Although this method samples a factor of $\sim$2 
different physical radii across our clusters (ranging from 2.2 Mpc for our
lowest redshift cluster, Abell~2199, to 5.2 Mpc for ZwCl~1215.1+0400),
the goal here is simply to provide an estimate of the redshift range to 
define a simple cluster membership criteria 
from -- within $3\sigma_{cz}$ of $\overline{cz}$.
An analogous approach is taken for Coma, but using a 2 degree radius (a 3.4 Mpc radius).
To place the clusters on to a common, physically meaningful scale, we limit our
subsequent analysis to those galaxies to within 
$r_{200} \approx R_{Virial} = 0.002\sigma_{cz}$ (Girardi et al.\ 1998), where
$r_{200}$ is the clustocentric radius at which the mean interior
density is 200 times the critical density; this value
is well approximated by $R_{virial}$. Although we could compute this radii in other ways
(e.g.\ Carlberg et al.\ 1997), we emphasize that 
this approximation is sufficient to serve to place
our clusters on to a common scale.
These values are tabulated in Table~\ref{tab:sample}.
Although it is known that
there is considerable scatter in the $L_X$-$\sigma_{cz}$ relationship
(Popesso et al.\ 2005),
the first conclusion to be drawn here is that Coma's 
velocity dispersion is not atypical compared to the
control sample (which has a mean of $1043\pm372$ kms$^{-1}$), but is is 
one of the largest given how we have selected the galaxy members.
We point out that the control sample has a full range of Bautz-Morgan (1970)
classifications (Table~\ref{tab:sample}) -- meaning we
cover a full range of galaxy cluster configurations and morphologies, 
ranging from those with obvious cD galaxies
centrally located in the clusters those lacking such a galaxy in entirety. Coma as 
a type II cluster that has two obvious, brightest cluster galaxies
is not atypical against this control sample: we do not regard it as more 
dynamically evolved than the control sample.

\section{X-ray Temperature}
%
%
\begin{figure*}
\centerline{\psfig{file=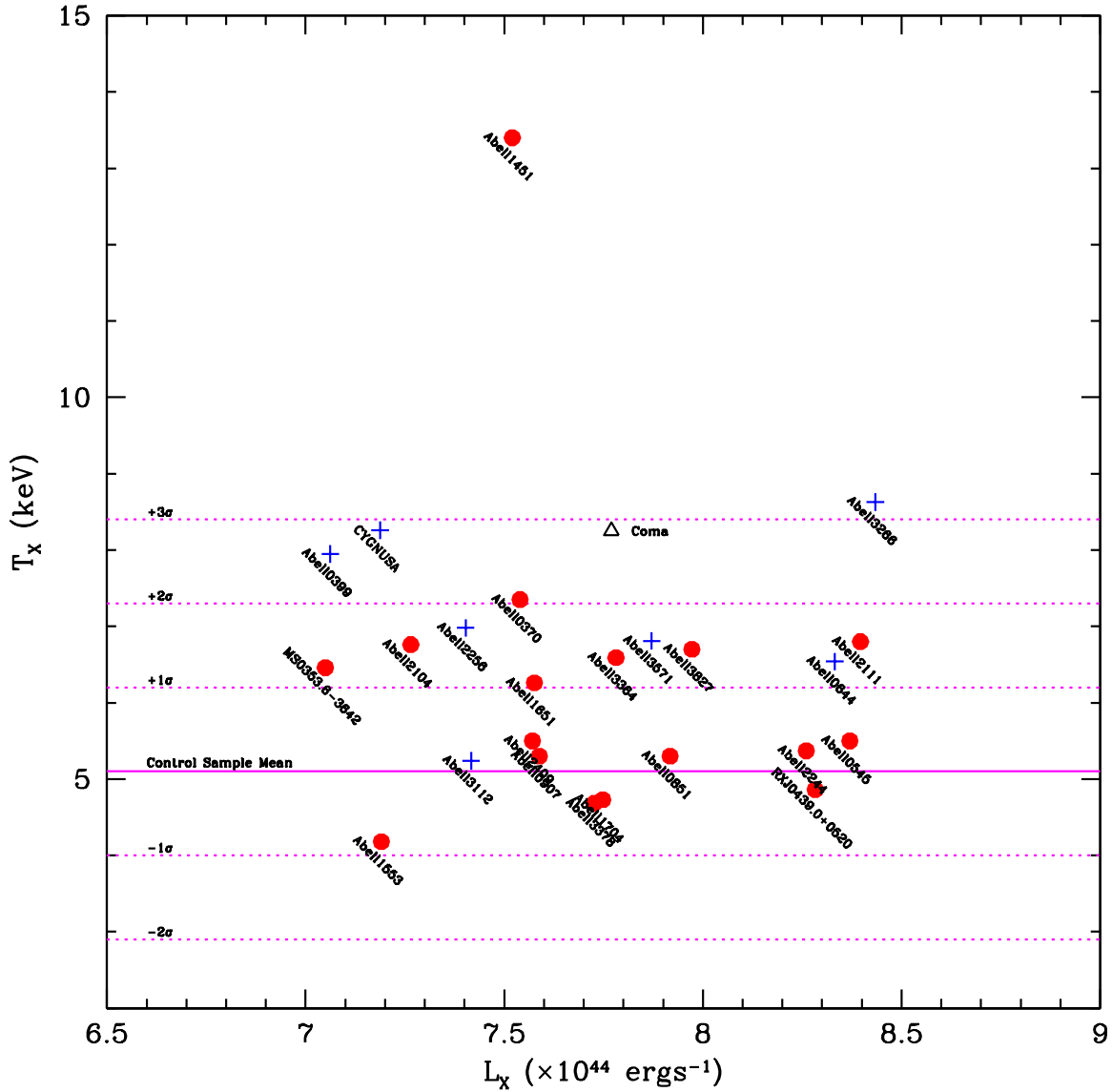,angle=0,width=6.5in}}
  \caption{X-ray temperatures for clusters extracted from
BAX within $1 \times 10^{44}$ ergs$^{-1}$ of Coma's
X-ray luminosity that have $T_X$ values available.
Clusters below $z=0.0814$ are marked with blue plusses,
those above with red filled circles.  Coma is marked by
a triangle near the centre of the plot.
The mean $T_X$ of our control sample (i.e.\ those clusters inside the SDSS
boundary that are within $5 \times 10^{44}$ ergs$^{-1}$ of Coma's
X-ray luminosity) is
denoted by the solid horizontal line and a few standard deviations
either side of this is represented by the dotted lines, as labelled.
Coma has one of the
largest $T_X$ values for this narrow $L_X$ range and is $\sim2.9\sigma$
above the control sample's mean $T_X$ value.
}
\label{fig:lxtx}
\end{figure*}

From Table~\ref{tab:sample}, it is already clear that Coma has the largest X-ray 
temperature (8.25 keV) out of all the comparable
clusters selected within SDSS. The mean temperature of our control sample
is $T_X = 5.1 \pm 1.1$ keV -- some $2.9 \sigma$ lower than the temperature
of Coma. Such a large temperature means that the physical conditions inside
Coma may actively regulate the star-formation of galaxies contained therein.
For example, Urquhart et al.\ (2010) notes that high $T_X$ clusters have a much
lower fraction of photometrically blue galaxies (i.e.\ Butcher-Oemler fraction;
Butcher \& Oemler 1984) than low $T_X$ clusters and are highly unlikely
to contain any extremely blue galaxies. Further, Popesso et al.\ (2007b)
and Aguerri et al.\ (2007) find an anti-correlation between $L_X$ and
cluster blue fraction which supports this finding given the scaling between
$L_X$ and $T_X$. This is reflected in the work of Poggianti et al.\ (2006)
who demonstrate a broad anti-correlation between cluster velocity dispersion
(a parameter that also scales with $L_X$; Dav{\'e} et al.\ 2002) and the fraction
of star-forming cluster galaxies.

Popesso et al.\ (2005) report the scaling relationship between $L_X$ and
$T_X$ in detail and show that there is both a trend and 
appreciable scatter between the two variables (see also Dav{\'e} et al.\ 2002).  
Although Coma's $T_X$ value may be significantly larger than our
control sample, we have used a factor of 2 range in $L_X$ to draw this conclusion
from.  To determine if its $T_X$ is truly anomolously high, we need to select 
clusters in a much narrower range of $L_X$.  We turn again to BAX to do this
and select \emph{all} available clusters within $1 \times 10^{44}$ ergs$^{-1}$ of Coma's
X-ray luminosity that also have a reliable X-ray temperature measurement available.  
In Fig.~\ref{fig:lxtx} we plot this narrow range of $L_X$ against $T_X$ for all 
available clusters.  Coma is again seen to have one of the highest temperatures
for all clusters in this range -- both above and below the redshift cut-off
of our control sample of $z=0.0814$.  But it is certainly within $2\sigma$ of the
mean $T_X$ of this narrower $L_X$ range sample.
That said, there is only one cluster either 
side of this redshift that has a larger X-ray temperature\footnote{We re-affirm the
note made by 
Valtchanov et al.\ (2002) about Abell~1451 ($T_X = 13.4$ keV; Matsumoto et al.\ 2001) 
possessing a very significant deviation
away from the $L_X$-$T_X$ scaling relation (e.g., Popesso et al.\ 2005).
This cluster merits future follow-up to discern the impact and potential cause of such 
an extreme temperature.}.

We therefore conclude that Coma's X-ray temperature is comparatively high: both against
our control sample, and against all available clusters in a much narrower $L_X$ range.

\section{Cluster sub-structure}
In this section, we address the second of our comparisons of Coma to the control sample
using global cluster kinematical approaches.
Depending on cosmological parameters such as 
the matter density of the Universe, it might be expected that a rich
cluster of galaxies (i.e.\ such as the ones that are in our sample) have perhaps 
had as much as 
half of their mass accreted within the past $\sim$few Gyr (e.g.\ Lacey \& Cole 1993).
Under such circumstance, it can be expected that a large fraction of
rich clusters exhibit measurable sub-clustering.  Coma is already well-known to
possess sub-clustering (see above).  But, what fraction of our control
sample also exhibits sub-clustering?  To determine this fraction, we
use the approach of Dressler \& Shectman (1988; DS) to evaluate if
the clusters possess significant sub-clustering. The test is powerful:
Pinkney et al.\ (1996) report that the DS approach is the most sensitive
test for sub-clustering from a swathe of tests that they evaluated.
The method works by computing a local mean local velocity ($\overline{cz}_{local}$)
and local velocity standard deviation ($\sigma_{local}$)
of a galaxy and its ten nearest neighbours.
These values are subsequently compared to the parent cluster's
mean velocity and $\sigma_z$ such that:
\begin{equation}
\delta^2 = \left( \frac{ N_{local} + 1 }{\sigma_v^2} \right)
[ (\overline{cz}_{local}-\overline{cz})^2 + ( \sigma_{local}-\sigma_v)^2 ]
\end{equation}
where $\delta$ is a measure of the deviation of the
individual galaxy.  
The parameter of merit, $\Delta$, is
computed as the summation of all $\delta$ terms.
This is contrasted to a Monte Carlo re-simulation of the
cluster where the galaxy velocities have been 
randomly shuffled
to each galaxy to generate $P(\Delta)$ and thereby 
estimate the confidence level that the cluster contains 
sub-structure. 

\begin{figure}
\centerline{\psfig{file=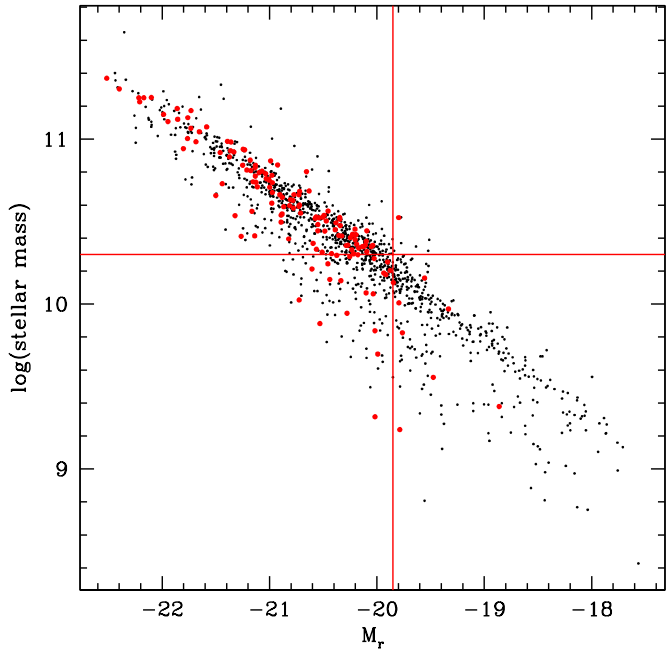,angle=0,width=3.65in}}
  \caption{Mass and absolute luminosity of all galaxies in the
control sample (smaller, black dots) with the contribution from
the most distant cluster in our sample, Abell~2255, overlaid
(larger, red dots). The SDSS limiting apparent magnitude of $r=17.77$
is transformed in to an absolute value using the mean redshift of
Abell~2255 and denoted by the vertical line. The mass limit of 
log(stellar mass)$=10.3$ (horizontal line) 
denotes the mass above which we are complete for the sample. 
We apply these two criteria to the entire control sample to ensure
we probe similar ranges in all clusters.
}
\label{fig:lims}
\end{figure}

Before we apply the DS test to our control sample, we need to
not only use the cluster membership criteria derived above and limit
the members to within $R_{Virial}$, but
also limit the cluster members to a similar absolute luminosity
range and mass range. This is necessary since substructure is 
strongly dependant on the galaxy luminosity range considered 
(Aguerri \& Sanchez-Janssen 2010). 
This is achieved by considering the highest 
redshift cluster in the control group: Abell~2255. For this cluster,
the SDSS limiting apparent magnitude of $r=17.77$ corresponds to an
absolute value of $-19.85$ (Fig.~\ref{fig:lims}). 
At this limit, we are mass complete
to log(stellar mass)$=10.3$ (Fig.~\ref{fig:lims}). 
We subsequently impose these two limits in absolute magnitude 
and mass on all of our cluster members.

%
%

Due to a paucity of data (less than 30 galaxy members per cluster) after
applying these cuts, we are forced to eliminate Abell~85, 660, 2199 and
Zwicky~1518.8+0747 from our control sample at this stage.
Of our sample, Abell~1775 and Abell~2065 (2 out of 8)
produce a $P(\Delta)$ statistic that is $<0.1$ per cent
(indicating certain substructure within $R_{Virial}$). We note that this
remains constant even if we ignore the absolute magnitude limit imposed above.
Given the 
comparatively large velocity dispersion of these clusters, 
this is perhaps expected (Hou et al.\ 2012).
Moreover, from $\Lambda$CDM simulations of clusters, Knebe \& Muller (2000) 
demonstrate that some 30 per cent of all clusters should exhibit subclustering
due to inter-cluster merger and infall activity (modulo a slightly different selection
criteria). We therefore regard Coma (and, indeed, our control group) 
as being "typical" for clusters in a $\Lambda$CDM Universe 
for the level of substructure observed at our limits.

\subsection{Velocity Dispersion Profiles}
%
%
\begin{figure*}
\centerline{\psfig{file=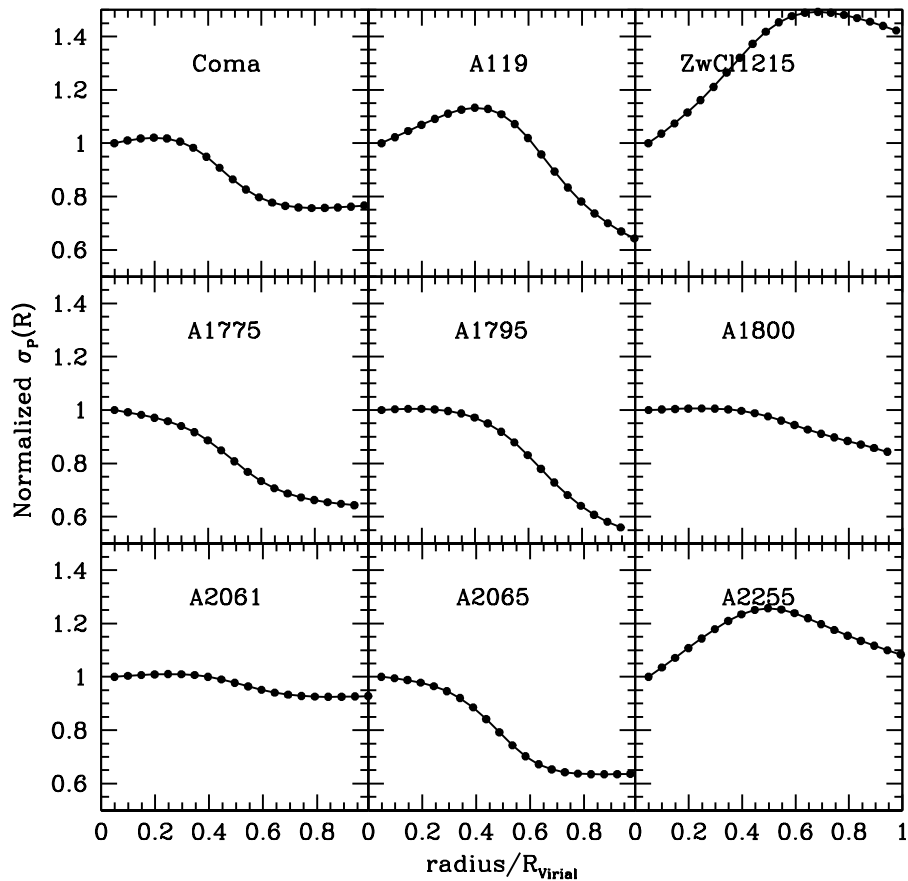,angle=0,width=5in}}
  \caption{Velocity dispersion profiles of our clusters. For each cluster,
$\sigma_P(R)$ is normalized to the central value. ZwCl1215 (top right) is arguably
the only cluster to display a strongly
rising profile with radius whereas the other clusters
either have a flat, falling, or combination profile.
}
\label{fig:vdp}
\end{figure*}

In recent years, a number of authors have probed how the velocity dispersion
profile of clusters is affected by various cluster-intrinsic factors such as
substructure (Hou et al.\ 2012) as well as potentially the 
dwarf to giant ratio (Pimbblet \& Jensen 2012) and the occupancy of 
the cluster by different spectral classes of galaxy (Rood et al.\ 1972). 
To complement the above analysis, we now compute the velocity
dispersion profile ($\sigma_P(R)$) of each of our clusters following the 
prescription of Bergond et al.\ (2006; see also Hou et al.\ 2012). Formally,

\begin{equation}
\sigma_P(R) = \sqrt{\frac{ \sum_i w_i(R) (x_i-\overline{x})^2 } {\sum_i w_i(R)} }
\end{equation}
where $x_i$ are the measured radial velocities of each galaxy and 
$\overline{x}$ is the mean recession velocity of the cluster taken from Table~1.
The weighting factors, $w_i$, are applied such that:
\begin{equation}
w_i(R) = \frac{1}{\sigma_R} exp \left( \frac{(R-R_i)^2}{2\sigma_R^2} \right)
\end{equation}
where $\sigma_R$, the kernel width, is a free parameter that we 
arbitrarily set to $0.2 R_{Virial}$. The velocity dispersion 
profiles computed in this manner are displayed in Fig.~\ref{fig:vdp}.
Interestingly, the clusters with significant sub-structure are not seen
to have a rising velocity dispersion profile. This argues that any local
kinematic group of galaxies may be at a late stage of homogenization with the wider
cluster.
This is in contrast to ZwCl~1215+0400 which does 
have a markedly rising profile and lack obvious substructure. This may be caused by
multiple sub-clumps at large radii infalling for the first time.
In comparison, Coma is quite un-remarkable set against these profiles.

\section{Star formation}
In this section, we determine the star-formation activity levels 
for cluster members in Coma and the control sample. 

\begin{figure}
\centerline{\psfig{file=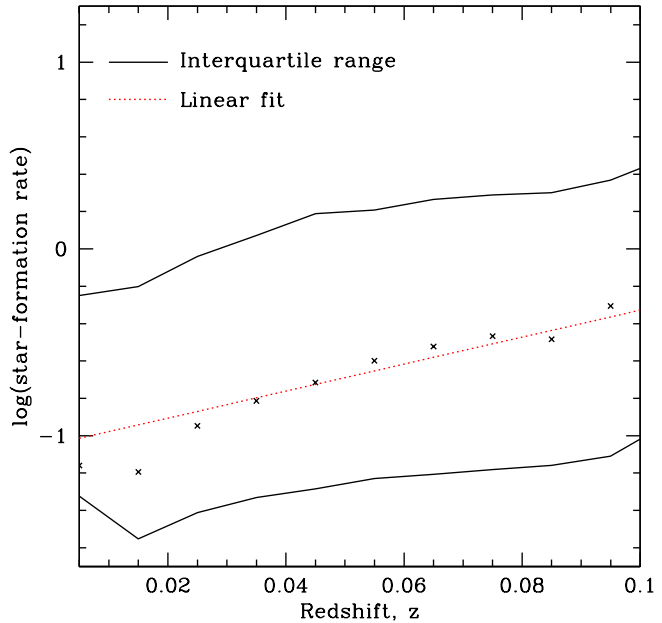,angle=0,width=3.5in}}
  \caption{Evolution of the galaxy main sequence for log(stellar mass)=10.4--10.6
SDSS galaxies up to $z=0.1$. The points are the median star-formation rates per redshift 
bin, whilst the
solid lines give the interquartile range of the distribution. As noted by Noeske et al.\ (2007),
the range of star-formation also evolves with $z$.
The linear fit (dotted line) to the data has a gradient of $7.22\pm0.21$ and we use
this fit to evolve all the data in the control sample to Coma's redshift with in the subsequent
analysis.
}
\label{fig:msevol}
\end{figure}

One way in which 
to do this is to use the galaxy main sequence (Noeske et al.\ 2007; and references therein):
a plot of star formation rate against galaxy mass.  This sequence is known
to evolve with redshift -- at high z, the average star formation rate of
galaxies is higher per galaxy mass than at lower z; the evolution in the
trend being largely attributed to gas exhaustion. Therefore, if we are to use
the galaxy main sequence to probe the activity levels in Coma and the control
sample, we must first correct for this redshift evolution.  We accomplish
this by accessing all SDSS galaxies in 0.005 redshift bins up to $z=0.1$ thereby
encompassing all of our sample. For each bin, we compute the median and inter-quartile
range of star-formation rates\footnote{Star-formation rates for the galaxies
are sourced from the SDSS value-added catalogue which are computed 
as per Brinchmann et al.\ (2004) using model fitting.}
of log(stellar mass)=10.4--10.6 galaxies (the choice
of this mass range is arbitrary, but is sufficiently representative of our own sample
and balances the needs to have good statistics to compute the redshift evolution of
the main sequence from). The results of this are displayed in Fig.~\ref{fig:msevol}.
We fit the data with a linear relationship which has a gradient of $7.22\pm0.21$ in
this range.  Although the actual evolutionary 
relation will likely by a higher order of $(1+z)$,
this linear relation is sufficient to describe these data at $z<0.1$.

\begin{figure}
\centerline{\psfig{file=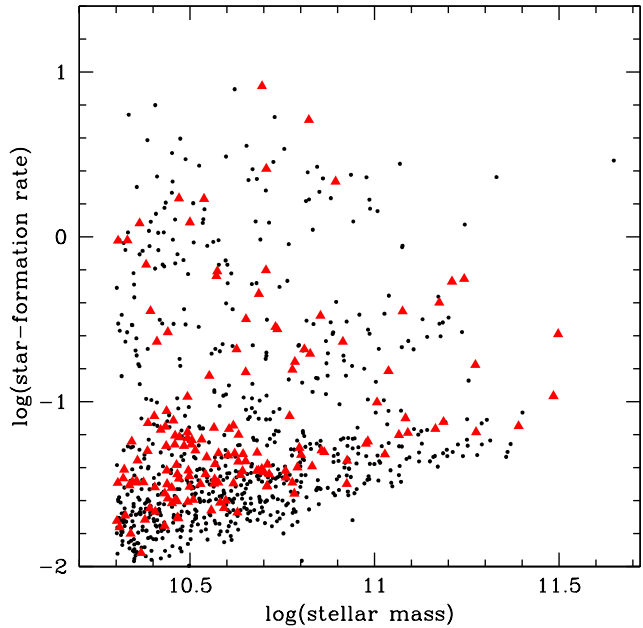,angle=0,width=3.5in}}
  \caption{Galaxy main sequence for Coma (red triangles) versus the
de-redshifted control sample (black dots). The galaxies in Coma have a
higher systematic average star formation rate at a given stellar mass than the control
sample.
}
\label{fig:mainseq}
\end{figure}

We use the gradient determined in Fig.~\ref{fig:msevol} to de-redshift the 
star-formation rates of galaxies in our control sample to that of Coma.
In Fig.~\ref{fig:mainseq}
we plot the galaxies from Coma and our de-redshifted control sample in the galaxy 
main sequence phase space (again, using 
data from the value added SDSS catalogue; Brinchmann et al.\ 2004). 
From this figure, we see that the galaxies in Coma appear to have a systematically
higher star formation rate at a given stellar mass than the control sample.

But is this apparent observation real? A two dimensional Kolmogorov-Smirnov (KS) test 
(Fasano \& Franceschini 1987; Peacock 1983) returns a very low chance
($<0.001$ per cent; coupled with a high $\Delta$ statistic) 
that the two distributions are drawn from the same sample. We therefore consider 
Coma to have a population whose galaxies possess significantly 
higher star formation rates on average
than comparable clusters at similar evolutionary stages. We visually inspect those
galaxies with very high star-formation and specific star-formation rates
and confirm that they appear to be late type (spiral and irregular) galaxies
that we assume are undergoing a starburst phase.

\begin{figure}
\centerline{\psfig{file=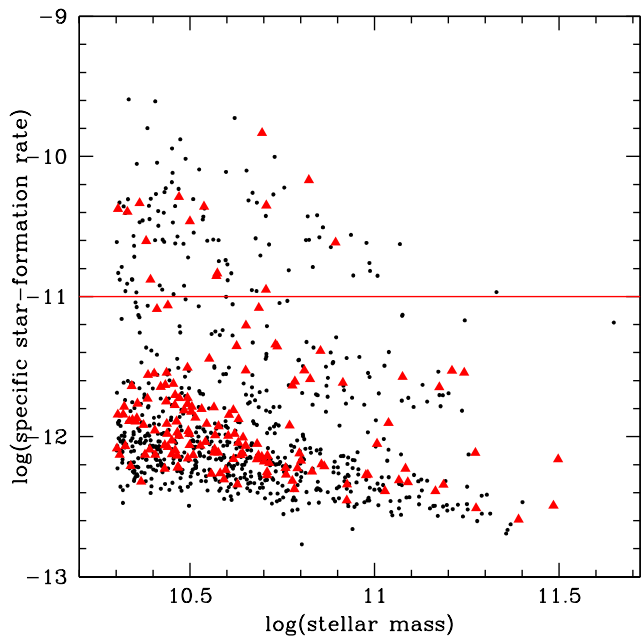,angle=0,width=3.5in}}
  \caption{Specific star-formation rates as a function of galaxy mass.
Symbols are the same as per Fig.~\ref{fig:mainseq}. The horizontal
line denotes the McGee et al.\ (2011) delimiter between active (above the line)
and passive (below the line) galaxies.  The fraction of active galaxies
differ between the two samples at a $\sim 2\sigma$ level.
}
\label{fig:ssfrm}
\end{figure}

A second way in which we may consider the active fraction is to use the divisor of
McGee et al.\ (2011) who use log(specific star formation rate)$=-11$ to differentiate
between active and passive galaxies. In Fig.~\ref{fig:ssfrm} we plot the specifc 
star-formation rate of Coma galaxies and the de-redshifted
control sample as a function of galaxy
mass. The fraction of galaxies that are active by this definition are $0.09\pm0.02$ 
in Coma versus $0.14\pm0.02$ for the control sample. This is $\sim 2\sigma$ (depending
on rounding) difference between the two samples. This appears to support (albeit at a weaker
level) the inference of the two dimensional KS test: the galaxies in Coma are systematically
different to the control sample.

%
%

\section{Discussion and Conclusions}

At the outset, we aimed to investigate three facets of Coma's galaxies in
comparison to a control sample: an X-ray temperature and luminosity comparison, 
a kinematic comparison, and a star formation activity comparison. One area
that we have deliberately avoided is an examination of the luminosity function of 
Coma. This is on the grounds that it has already
been well-studied in comparison to other
clusters at multiple wavelengths 
(recent examples include but are not limited to: Yamanoi et al.\ 2012;
Bai et al.\ 2009;
Cortese et al.\ 2008;
Adami et al.\ 2007) 
and will likely follow the kinematic results for the control 
sample (in the sense that multiple components may be reflected in a
superposition of functions; see also Tempel et al.\ 2009).

One way in which the situation of higher $T_X$ coupled with higher star-formation 
rate per galaxy mass bin being larger in Coma compared to the control sample might be 
arranged is if the bluer galaxies in Coma are just arriving in to the cluster environment
(given that a hotter intra-cluster medium should inhibit galaxy star formation subsequently).
This ties with Mahajan et al.\ (2012)'s finding: a high
galaxy density in the infalling and filamentary regions of clusters such as Coma inevitably leads to
a greater rate of galaxy-galaxy interaction and consequentially an increased starburst rate.
But the problem with this interpretation is that there are $\sim$equally massive clusters
in the control sample by design (i.e.\ the X-ray selection used here).

There have been hints in the literature that some of the special features
of Coma might be allieviated if the distance to Coma was slightly lower.
Consider for example Fig.~6 of van Dokkum \& van der Marel (2007) which 
shows that the mass to light ratio of Coma is similar to that of $z\sim0.2$ clusters.
If the distance to Coma were lower, then this ratio would increase, bringing 
Coma's mass to light ratio more in line with the trend observed with redshift
by the same authors.  This could be achieved if Coma had a none-negligible 
peculiar velocity with respect to the Hubble flow (e.g., toward the Shapley concentration). 
An interesting facet of this hypothetical change
would be a driving of the star-formation rates of Coma galaxies lower -- bringing
them more in line with the de-redshifted control sample points.
Given results
that suggest Coma has been reported to have 
negligible peculiar velocity (e.g., Bernardi et al.\ 2002)
and a variety of measurements agreeing within 
uncertainty on its distance (e.g., Capaccioli et al.\ 1990; 
D'Onofrio et al.\ 1997; Jensen et al.\ 1999;
Kavelaars et al.\ 2000;
Liu \& Graham 2001),
we do not view this as a likely scenario; 
we supply it simply as an illustration.

\ \\ 
In summary, in this work we have shown:

(i) Although Coma has a large velocity dispersion, it is not atypical for 
a cluster of its $L_X$. However, the X-ray temperature of Coma is rather high:
some $2.9\sigma$ hotter than our control sample. Even considering all clusters
available with a published $T_X$ within $1\times10^{44}$ ergs$^{-1}$ of Coma
reveals it has one of the highest temperatures for all clusters in the range.
Given the relationship between $T_X$ and cluster galaxy properties we urge
strong caution in using Coma as a $z\sim0$ baseline for studying cluster
galaxy evolution.

(ii) Coma is well-known to contain sub-structure. In comparison, 
we show 2 out of 8 cluster in the control sample also contains significant
sub-structure within $R_{Virial}$. Coma is therefore un-remarkable in this
regard.

(iii) The velocity dispersion profiles of the control sample contain a mixture
of rising, falling, flat and combination profiles.  Coma is un-remarkable set
against this background and reinforces the above conclusion that
Coma is kinematically normative for clusters of its ilk. 

(iv) The general star-formation rate of Coma cluster galaxies 
inferred from the galaxy main sequence is systematically higher
than for the control sample. A two dimensional KS test rejects the hypothesis
that the two samples are drawn from the same parent population with over 99 per cent
confidence. Further, the fraction of actively star forming galaxies by the definition of
McGee et al.\ (2011) is $0.09\pm0.02$ for Coma, versus $0.14\pm0.02$ for the 
control sample. We note in speculation that this discrepency could be alleviated
if the distance to Coma were smaller.

Thus, whilst Coma might be kinematially ``typical'', the galaxies contained within
are less suppressed in star-formation rate than the comparison clusters. 
We consequentially urge caution in using Coma
as a $z\sim0$ cluster in galaxy evolution works: its galaxy population to the
limits probed by this sample are not typical of clusters for its mass 
(as approximated by its X-ray luminosity).

%
%

%
%

\section*{Acknowledgements}
KAP thanks Christ Church College, Oxford,
for their hospitality whilst the bulk of this work was being undertaken.
SJP is a Super Science Fellow at Monash University. KAP and SJP thank 
the Australian Research Council for their support through 
grant number FS110200047.  We thank Ryan Houghton and 
Martin Bureau for valuable 
discussion during the preparation of this work.

We would like to express our gratitude to the anonymous 
referee for her/his robust feedback on the earlier versions
of this manuscript.

This research has made use of the X-Ray Clusters Database (BAX)
which is operated by the Laboratoire d'Astrophysique de Tarbes-Toulouse (LATT),
under contract with the Centre National d'Etudes Spatiales (CNES) 

Funding for the SDSS and SDSS-II has been provided by the Alfred P. Sloan Foundation, the Participating Institutions, the National Science Foundation, the U.S. Department of Energy, the National Aeronautics and Space Administration, the Japanese Monbukagakusho, the Max Planck Society, and the Higher Education Funding Council for England. The SDSS Web Site is http://www.sdss.org/.

The SDSS is managed by the Astrophysical Research Consortium for the Participating Institutions. The Participating Institutions are the American Museum of Natural History, Astrophysical Institute Potsdam, University of Basel, University of Cambridge, Case Western Reserve University, University of Chicago, Drexel University, Fermilab, the Institute for Advanced Study, the Japan Participation Group, Johns Hopkins University, the Joint Institute for Nuclear Astrophysics, the Kavli Institute for Particle Astrophysics and Cosmology, the Korean Scientist Group, the Chinese Academy of Sciences (LAMOST), Los Alamos National Laboratory, the Max-Planck-Institute for Astronomy (MPIA), the Max-Planck-Institute for Astrophysics (MPA), New Mexico State University, Ohio State University, University of Pittsburgh, University of Portsmouth, Princeton University, the United States Naval Observatory, and the University of Washington.

This research has made use of the NASA/IPAC Extragalactic Database (NED) which is operated by the Jet Propulsion Laboratory, California Institute of Technology, under contract with the National Aeronautics and Space Administration.

\end{document}